\documentstyle[epsf]{mn}

\title[Serendipitous X-ray flares]{Properties of serendipitous
X-ray flares discovered in $XMM-Newton$ observations}

\author[Trenholme, Ramsay \& Foley]{
Duncan Trenholme, Gavin Ramsay, Carl Foley\\
Mullard Space Science Laboratory, University College London,
Holmbury St. Mary, Dorking, Surrey, RH5 6NT, UK\\}
\date{Received: }

\begin{document}
\outer\def\gtae {$\buildrel {\lower3pt\hbox{$>$}} \over 
{\lower2pt\hbox{$\sim$}} $}
\outer\def\ltae {$\buildrel {\lower3pt\hbox{$<$}} \over 
{\lower2pt\hbox{$\sim$}} $}
\newcommand{\ergscm} {ergs s$^{-1}$ cm$^{-2}$}
\newcommand{\ergss} {ergs s$^{-1}$}
\newcommand{\ergsd} {ergs s$^{-1}$ $d^{2}_{100}$}
\newcommand{\pcmsq} {cm$^{-2}$}
\newcommand{\ros} {\sl ROSAT}
\newcommand{\exo} {\sl EXOSAT}
\newcommand{\chandra} {\sl Chandra}
\newcommand{\xmm} {\sl XMM-Newton}
\def\rchi{{${\chi}_{\nu}^{2}$}}
\newcommand{\Msun} {$M_{\odot}$}
\newcommand{\Mwd} {$M_{wd}$}
\newcommand{\Zsun} {$Z_{\odot}$}
\newcommand{\Rsun} {$R_{\odot}$}
\def\Mdot{\hbox{$\dot M$}}
\def\mdot{\hbox{$\dot m$}}

\maketitle

\begin{abstract}

We present the results of a search of the {\xmm} public data archive
for stellar X-ray flares. We find eight flaring sources for which we
identify 7 optical counterparts. Three of these sources have distance
estimates which allow us to determine their luminosities. Based on the
decay time of the flares and their luminosity we derive loop
half-lengths of $\sim 2-7\times10^{10}$cm and emission measures of
$\sim 10^{54}$cm$^{-3}$: these are similar to values derived for other
stellar flaring sources. One of the stars shows two flares in close
succession. We discuss the likelihood of this double event being
either sympathetic or homologous in nature. A comparison to a pair of
similar flares on the Sun suggests that homology is the more likely
process driving the flare event.

\end{abstract}


\section{Introduction}

Solar and stellar flares have been studied using both ground and space
observatories. Solar observations have revealed that the energy source
which powers Solar flares most probably originates with the
destruction and reconnection of Solar magnetic fields, as they relax
to a lower energy configuration (Priest \& Forbes 1999). The excess of
energy which was stored in the stressed pre-flare Solar magnetic field
is released in an outburst over most of the electromagnetic spectrum,
from Radio through to X-ray.

Flares which have been observed on stellar sources, are substantially
more powerful than their Solar counterparts by many orders of
magnitude (eg Stern, Underwood \& Antiochos 1983). These include
T-Tauri stars where the young rapidly rotating protostar has a
relatively high magnetic field. The magnitude and frequency of the
flares typically reduce as the protostar enters the main
sequence. This has been attributed to the reduction in their rotation
rate, which results in a slowing of the stars internal dynamo, the
process responsible for generation of stellar magnetic fields.

Flare stars are usually discovered as a result of intensity
variability studies. One systematic search was that of the {\ros} all
sky survey which revealed 767 X-ray variable stars many of which where
found to be flare stars with both long duration flare activity and
sporadic short flare activity (Fuhrmeister \& Schmitt 2003). Whilst
this survey was unbiased, it was impossible to study the temporal
variation of the flares in any detail since the duration of each
observation was short.

In contrast, the longer orbit of {\xmm} allows uninterrupted
observation of sources for many hours. Further, {\xmm} has the largest
effective area of any imaging X-ray satellite, and in comparison with
{\ros}, has a much broader energy band and a higher spectral
resolution. A search of the {\xmm} data archive provides an excellent
opportunity to conduct a search for new sources which are flaring in
X-rays. Each flare can be studied in detail and will add to the
increasing database on X-ray stellar activity in general.

In this paper we present the results of our search for stellar flares
using the serendipitous {\xmm} data held in the public archive. We
show the results of our analysis of these flares which include an
estimate of their luminosity and the nature of the source from which
it originated. We estimate the loop half-length of four flares, and
compare our results with that of other stellar flares. We also
investigate the process which may have caused a double flare event
which was observed on one of the sources and then compare its
properties with Solar flare events.

\begin{table}
\begin{center}
\begin{tabular}{lrcr}
\hline
Date & Orbit & ObsId & Clean Time\\
     &     &       &  (ksec) \\
\hline 
2002-02-20 & 0403 & 0007422401 & 9.0\\                         
2002-03-24 & 0419 & 0112640201 & 21.7\\
2002-06-15 & 0461 & 0041750101 & 51.6\\
2002-08-13 & 0490 & 0101440201 & 50.4\\
2002-09-15 & 0507 & 0112530101 & 34.0\\
2002-09-15 & 0507 & 0049560301 & 13.8\\
2002-12-16 & 0553 & 0148440101 & 25.0\\
\hline
\end{tabular}
\end{center}
\caption{The observation log for the selected {\xmm} observations. We
show the date of the observation, the {\xmm} orbit number, the
observation id and its clean time duration.}
\label{obs}
\end{table}

\begin{figure*}
\begin{center}
\setlength{\unitlength}{1cm}
\begin{picture}(13,10.7)
\put(-1.5,0){\includegraphics{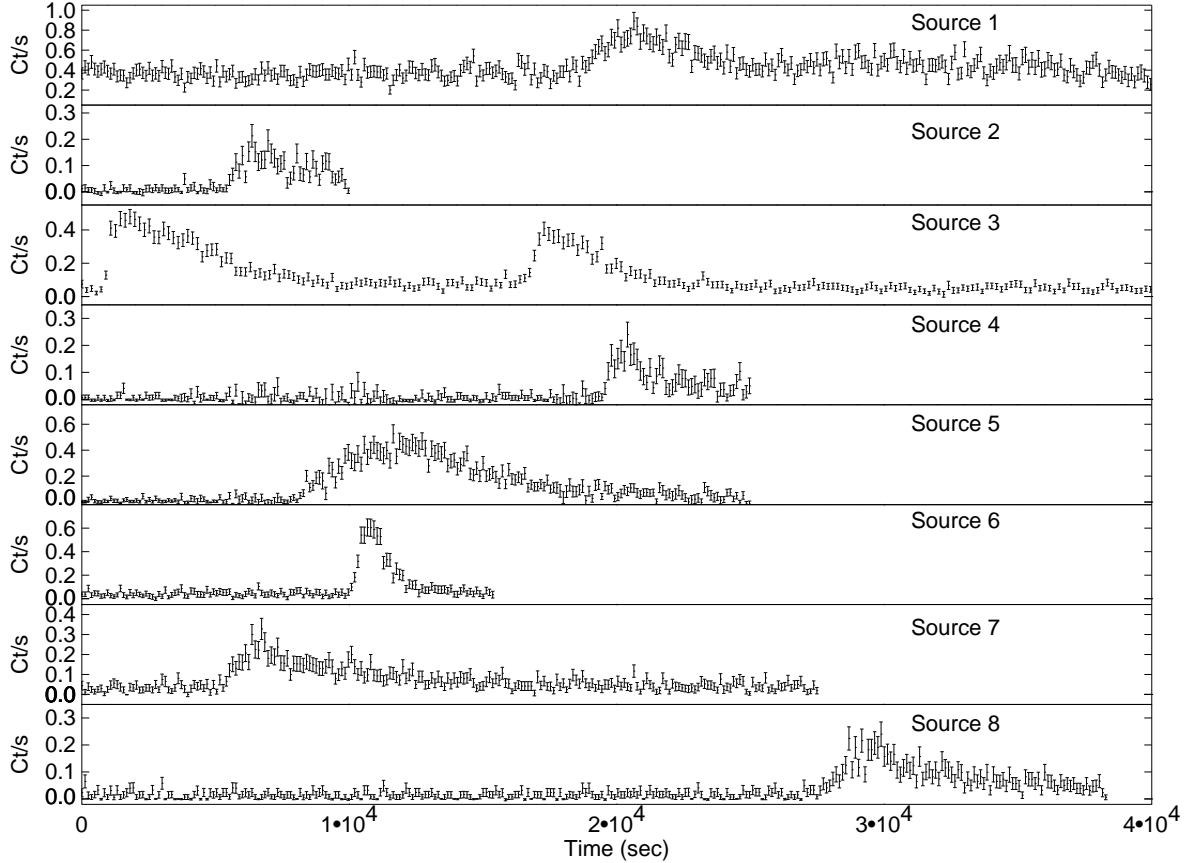}}
\end{picture}
\end{center}
\caption{The background subtracted X-ray light curves for each of the
8 flares in the energy range 0.2-5.0keV. All light curves are extracted
from the EPIC pn instrument, apart from Source 3, which is from the
EPIC MOS 1+2 instrument. The bin size is 120 sec for all except Source 3
which is 180 sec}
\label{light}
\end{figure*}

\section{Observations}

{\xmm} was launched in Dec 1999 by the European Space Agency (Jansen
et al. 2001). The EPIC instruments have imaging detectors covering the
energy range $\sim$0.15-10keV with moderate spectral resolution. Since
our main interest is searching for X-ray flares: we therefore use data
primarily taken using the EPIC pn detector since this has the highest
effective area of the imaging X-ray detectors (Str\"{u}der et
al. 2001). All the observations for this search were selected from the
{\xmm} public data archive. The selected observations were chosen with
bias toward fields that potentially contained a large number of X-ray
sources, and bias against fields that contained clusters of galaxies,
since they were less likely to contain flaring sources (for brevity we
do not note the individual fields which were selected). In this paper
we present the results of our analysis of the first 100 observations
we selected from the {\xmm} public data archive.

The data were processed using the {\xmm} Science Analysis Software
(SAS) v5.3.3. Each field was source searched and the light curve of
each detected source was tested for variability (see Ramsay, Harra \&
Kay 2003 for details of this procedure). A total of 8 sources were
found to show clear flare-like variability, ie a rapid (less than a
few ksec) increase in the X-ray count rate followed by a decline
lasting less than $\sim$10ksec. For these sources we obtained
background-subtracted light curves, typically with a bin-size of 120
sec. Table \ref{obs} shows the {\xmm} observation logs for the fields
which contained these flares. Other sources were found to be variable,
but did not show flare-like behaviour.

Using the SIMBAD Astronomical database and the USNO-B1.0 optical
catalogue, we searched for optical counterparts of our X-ray flaring
sources.  For seven of our eight sources we found a counterpart (Table
\ref{sources}). The offset between the {\xmm} and optical counterpart
was typically less than 3 arcsec. This is significantly less than the
FWHM of sources in the EPIC pn. We note in Table \ref{sources} the
USNO catalogue number of each source, the $B$ and $R$ magnitudes from
two epochs, and the $V$ and $B$ magnitudes from SIMBAD. There is
evidence that some of these sources show significant variation in
their optical brightness between epochs. We have not identified a
optical counterpart to Source 8, although it is located in the
immediate vicinity of $\zeta$ Ori, which may have prevented optical
surveys reaching as deep as usual.

\begin{table*}
\begin{tabular}{rccrccccccc}
\hline
 Source & RA & Dec & USNO-B1.0 &
\multicolumn{6}{c} {Magnitude} & XMM\\
 & \multicolumn{2}{c} {(2000.0)} & & $V$ & $(B-V)$ & $B_1$ & $B_2$ & $R_1$ & $R_2$ & Obs ID\\
\hline
1 & 10 42 41.5 & -64 21 05 & 0256-0180983 & 10.62 & 0.64   & 10.50 & 11.09 & 11.45 & 11.47 & 0101440201\\
2 & 13 57 03.4 & -58 47 10 & 0312-0398114 &       &        &       & 16.92 & 14.92 & 15.28 & 0007422401\\
3 & 00 02 56.4 & -30 04 46 & 0599-0000714 & 12.13 & 0.85   & 13.54 & 13.00 & 11.31 & 12.15 & 0041750101\\
4 & 05 42 12.3 & -02 05 10 & 0879-0104649 & 15.7  &        & 16.58 & 15.58 & 14.95 & 14.64 & 0112640201\\
5 & 05 41 45.5 & -02 24 17 & 0875-0105863 &       &        &       & 21.13 & 19.14 & 19.40 & 0112640201\\ 
6 & 05 35 22.3 & -04 25 28 & 0855-0060617 & 6.25 & -0.15 & 5.96  & 6.14  & 6.30  & 6.32  & 0049560301\\
7 & 01 40 31.3 & -68 05 09 & 0219-0011450 &       &        &       & 14.69 & 12.74 & 12.71 & 0148440101\\
8 & 05 40 52.5 & -02 00 48 &              &       &        &       &       &       &       & 0112530101\\
\hline
\end{tabular}
\caption{The X-ray sources which showed significant X-ray variability
in our survey. The X-ray position is the position from the {\xmm} EPIC
pn image.  The $V$ and ($B-V$) mag are taken from SIMBAD, while $B$
and $R$ magnitudes are from the USNO-B1.0 catalogue at 2 epochs. We
also indicate the {\xmm} observation number (cf Table 1) in which the source
was detected.}
\label{sources}
\end{table*}

\begin{figure}
\begin{center}
\setlength{\unitlength}{1cm}
\begin{picture}(11,10)
\put(0,-2.5){\includegraphics{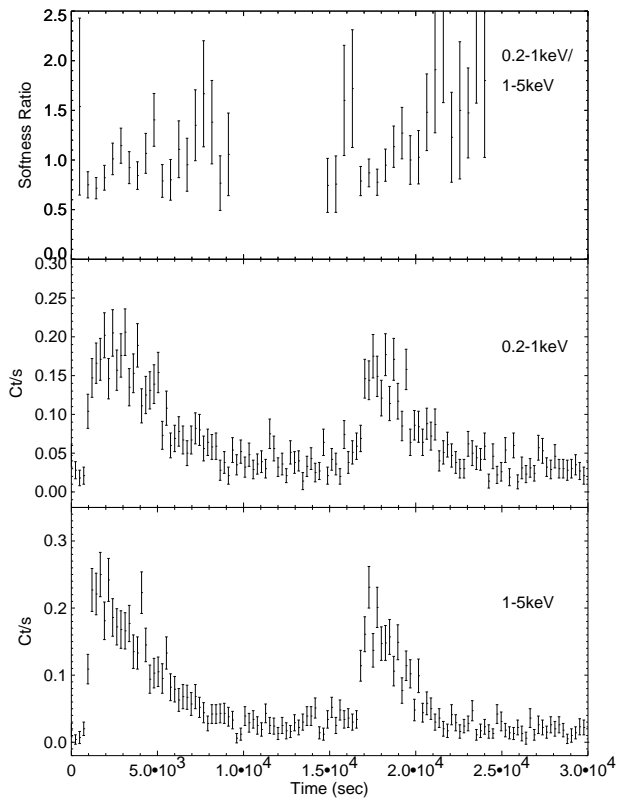}}
\end{picture}
\end{center}
\caption{The double flare seen on Source 3 in the 0.2-1 keV
(middle panel) and 1-5 keV (bottom panel) and the 0.2-1/1-5 keV
softness ratio (top panel).}
\label{soft} 
\end{figure}

\section{Light curves}

In Figure 1, we present the background-subtracted X-ray light curves
for each flaring source we found. The light curves resemble those of
typical Solar flares, some being impulsive (Sources 2, 3, 4 and 6),
long duration (Source 5) and intermediate (Sources 1, 7 and 8) in
nature.

The sources with impulsive flares all show a rise to their maximum
brightness on a timescale of $\sim$500 sec. The flare on Source 6 is
particularly short, lasting less than $\sim$1 hr (the time duration
for which flux is observed to be above the normal quiescent level),
while for Source 2 and 4 the observation finishes before the end of
the flare. In comparison Sources 7 and 8 show a rise time
approximately 4 times as long as the rapid flares, with the decay time
also being relatively long, indeed the decay continues beyond the end
of the observation.

In contrast, in Source 5, the rise to maximum is much more gradual
than the sources already discussed, and the maximum lasts much
longer. Source 1 differs to the other sources in that a significant
flux is seen before the increase in flux, with the light curve profile
being broadly similar in shape to Source 5.

Perhaps the most interesting of the sources is the third, which
displays two impulsive flares of similar size and duration, separated
by $\sim$4 hours 20 mins. (The EPIC MOS 1+2 light curve has a longer
time duration than the EPIC pn light curve, so the MOS 1+2 light curve
is shown in Figure 1). The first flare lasts for $\sim$150 mins while
the second flare lasts for a slightly shorter time, with a duration of
$\sim$120 mins. (This flaring source is briefly alluded to by
Pillitteri et al. 2004 in an analysis of the same {\xmm} dataset).

To investigate this double flare in more detail we extracted a
background-subtracted light curve in the 0.2--1 and 1--5 keV energy
bands. We show these light curves and the resulting softness curve in
Figure \ref{soft}. The softness ratio curve shows evidence for a
softening over the duration of both flares. This is consistent with
the flare observed from JTP 138 in NGC 2516 (Ramsay et al. 2003).

\section{Source Spectra}
\label{Spec}

We extracted spectra for each of the 9 flares found on the 8 sources,
using the time interval of the flare duration only. For each flare a
background spectrum was also extracted from a nearby source-free area
of the detector.  Since the sources were not on-axis we generated
response and auxiliary files for each source using the SAS tasks {\tt
rmfgen} and {\tt arfgen}. Further, we used only good events ({\tt
FLAG=0}) and single and double events ({\tt PATTERN=0-4}). The
spectrum was then fitted using the X-ray package {\tt XSPEC} (Dorman
\& Arnaud 2001). 

We initially fit all spectra using an absorbed thermal plasma
model. All spectra required the addition of a second thermal plasma
component to achieve good fits. We show the results of the spectral
fits for all the flares in Table \ref{fits}. We note that these
results may be affected as a result of spectral softening over the
duration of the stellar flares (c.f Figure \ref{soft}). We also show
the observed flux in the 0.1--10keV band and for those sources for
which we know the distance (\S \ref{nature}) we show the bolometric
luminosity for the flare. We find the low temperature plasma has a
temperature $kT\sim$0.7keV, and a range of temperature for the hotter
plasma component. In some sources, the fits were significantly
improved when we allowed the metal abundance of the plasma to vary
from Solar (Table \ref{fits}). There is no relationship between metal
abundance and observed flux, suggesting that this effect is not simply
related to the signal-to-noise ratio of the spectrum.

\begin{table*}
\begin{tabular}{rccccrrrr}
\hline Source & $N_{H}$ & $kT_{1}$ & $kT_{2}$ & Abundance & Observed flux &
\multicolumn{2}{c} {Bolometric luminosity} &\rchi\\ & $\times10^{20}$ 
\pcmsq & (keV) & (keV) & \Zsun  &\ergscm & 10$^{30}$ \ergss & 10$^{34}$ ergs &
(dof)\\

\hline 1 & 10$^{+2}_{-2}$ & 0.7$^{+0.0}_{-0.1}$ & 7.8$^{+9.9}_{-3.2}$
& 0.07 &1.52$^{+0.42}_{-0.36}\times10^{-12}$ & 9.3$^{+2.6}_{-2.2}$ &
7.5$^{+2.0}_{-1.8}$ & 1.18 (116)\\

2 & 0.2$^{+1.8}_{-0.2}$ & 0.7$^{+0.4}_{-0.3}$ & 2.8$^{+1.2}_{-0.6}$ &
1 &5.38$^{+1.22}_{-0.94}\times10^{-13}$ & & & 0.57 (11)\\

3a & 0.7$^{+1.0}_{-0.7}$ & 0.7$^{+0.0}_{-0.1}$ & 2.1$^{+0.3}_{-0.2}$ &
0.34 &8.35$^{+1.82}_{-1.57}\times10^{-13}$ & 8.8$^{+1.9}_{-1.6}$ &
9.7$^{+2.1}_{-1.8}$ & 1.09 (95)\\

3b & 1.8$^{+1.4}_{-1.6}$ & 0.7$^{+0.1}_{-0.1}$ & 1.7$^{+0.5}_{-0.3}$ &
0.15 & 5.14$^{+2.92}_{-2.57}\times10^{-13}$ & 6.2$^{+4.2}_{-3.0}$ &
5.6$^{+3.8}_{-2.7}$ & 0.82 (55)\\

4 & 8.3$^{+4.9}_{-3.8}$ & 0.9$^{+0.1}_{-0.2}$ & $\geq$4.5 & 0.08 &
6.90$^{+3.83}_{-4.15}\times10^{-13}$ & & & 1.21 (18)\\

5 & 5.2$^{+1.1}_{-1.0}$ & 0.7$^{+0.1}_{-0.4}$ & 5.0$^{+0.7}_{-1.0}$ &
1 & 2.32$^{+0.07}_{-0.18}\times10^{-12}$ & & & 1.23 (101)\\

6 & 3.3$^{+2.5}_{-1.9}$ & 0.6$^{+0.1}_{-0.2}$ & 5.2$^{+1.7}_{-1.2}$& 1
& 1.83$^{+0.22}_{-0.21}\times10^{-12}$ & 34.3$^{+4.3}_{-4.0}$ &
17.5$^{+2.1}_{-2.1}$ & 1.23 (29)\\

7 & 0.7$^{+1.3}_{-0.7}$ & 0.9$^{+0.1}_{-0.1}$ & $\geq$1.95 & 0.08 &
6.04$^{+3.35}_{-3.17}\times10^{-13}$ & & & 1.54 (39)\\

8 & 3.9$^{+4.0}_{-2.4}$ & 0.4$^{+0.2}_{-0.1}$ & 3.3$^{+0.1}_{-0.1}$& 1
& 4.26$^{+0.72}_{-0.70}\times10^{-13}$ & & & 1.14 (28)\\

\hline
\end{tabular}
\caption{The spectral fits to the 9 X-ray flares. We have
extracted events covering only flare time interval.}
\label{fits}
\end{table*}

We show the spectrum of the brightest flare (Source 5) in Fig. 3: the
fit is good with no evidence of significant residuals at particular
energies.  This source was also the only system to show evidence for
an Fe line at 6.6$^{+0.3}_{-0.2}$keV: this may simply be due to the
fact that the other spectra have lower signal-to-noise ratios,
especially at higher energies. We note that the brightest flaring
source in NGC 2516 (JTP 138) also showed an Fe line consistent with
this line energy (Ramsay et al. 2003).

We show the average flare luminosity for 4 flares in Table
\ref{fits}. These range between 6--34$\times10^{30}$ \ergss. The most
luminous flare is Source 6, although since it is of short duration the
total energy released is comparable to the less luminous
flares. However, at its brightest it is an order of magnitude more
luminous than the flare source JTP 138 in NGC 2516.

\begin{figure}
\begin{center}
\setlength{\unitlength}{1cm}
\begin{picture}(7,5.5)
\put(-1.2,-0.8){\includegraphics{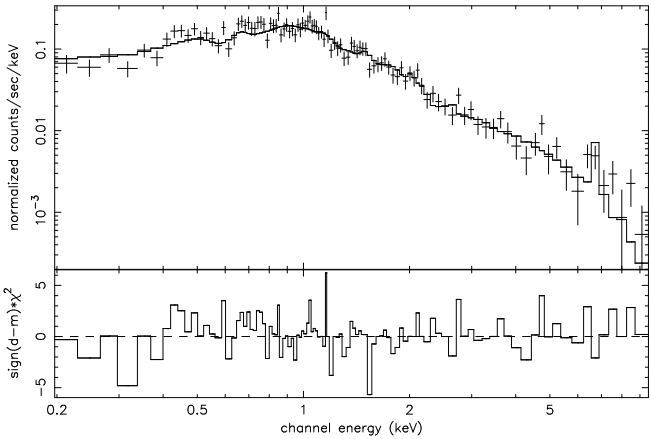}}
\end{picture}
\end{center}
\caption{The EPIC pn spectrum of Source 5. The best fit (shown as a
solid line) is an absorbed two temperature plasma model.}
\label{spec} 
\end{figure}

\section{The nature of the sources}
\label{nature}

We have identified seven of the eight sources as having an optical
counterpart. Of these, 3 are situated in the same field as known open
clusters.

Source 3 is in the same field as the open cluster Blanco 1. To
determine the likelihood of Source 3 being a member of this cluster we
compared the $V$, $B-V$ data of the source with the locus of the main
sequence (de Epstein \& Epstein 1985) of Blanco 1: we conclude that
this source is very likely a cluster member. For the calculation of
the luminosity of the flares from Source 3, a distance of 262 pc was
used (the distance to Blanco 1, Robichon et al. 1999). Panagi et
al. (1994) found ambiguous optical spectroscopic evidence for binarity
in this source.

Four other sources in this cluster were flagged as being X-ray
variable with a confidence of $>$99\% (Micela et
al. 1999). Lightcurves were extracted for the three sources that were
found within the {\xmm} field of view. Two of these sources were found
to significantly vary within the observation, both between 0.1 and 0.5
ct/sec, but they did not show flare-like behaviour.

Source 6 is identified with the bright star HD37016. It has a distance
of 346 pc based on the HIPPARCOS catalogue (Perryman et al. 1997). It
has a B2.5V spectral type and is a member of the cluster NGC 1977.

Source 1 is the only other of the eight sources to be classified
spectroscopically. Braes (1962) found this source to have a G0
spectral type. Barnes et al. (1999) identify this source as a variable
star with period 0.57 $\pm0.01$ days: it was also noted that the
optical lightcurves suggest the differential rotation of migrating
starspots. The source is located in the field of cluster IC
2602. Based on comparisons with $V$, $R-V$ data of the main sequence
(Prosser, Randich \& Stauffer 1996) we conclude this source is very
likely to be a member of this cluster. The source was also found to
have a radial velocity consistent with membership (Randich et al.
1997). The distance to cluster IC 2602 is 150 pc (Prosser et al.
1996) which we used in the calculation of the luminosity of the flare
event for this source.

Source 4 is an emission line star, with $V\sim$15.7, which was found
to be near nebulosity (Wiramihardja et al. 1989). There is another
optical source identified within one arcsecond of this source and is
highly reddened and much fainter ($V\sim$20) compared to the
counterpart identified in Table \ref{sources}. The two sources cannot
be resolved on the Digital Sky Survey image. Source 7 was found to be
an X-ray source in the Wide Angle {\ros} Pointed Survey (Perlman et
al.  2002).

Various systems exhibit flare like behaviour similar to the sort we
have observed in the sources presented here. These include late main
sequence stars; Young Stellar Objects (YSOs); RS CVn binary systems;
single early type stars; and binaries containing M dwarf stars. Single
late type main sequence stars usually show flare luminosities orders 
of magnitude less than those reported here (e.g. Tsikoudi, Kellett, \&
Schmitt 2000). Although YSOs show flare luminosities and intensity
profiles consistent with those three sources for which we have
distance estimates, Sources 1 and 3 are located in open clusters in
which star formation has ceased and are therefore unlikely to be
YSOs. In contrast Source 6 is located in the Orion star forming
region, indicating that it may be a YSO. RS CVn systems, binary
systems containing an M dwarf, or single early type stars, have flare
luminosities similar to that observed for our sources. To settle the
nature of these objects, a short series of medium resolution spectra
is necessary.

\section{Flare analysis}

Soft X-ray observations have revealed that the Sun consists of many
hot X-ray emitting coronal loops which trace the coronal magnetic
field.  Flares occur when these loop systems interact by a process of
magnetic reconnection (Priest \& Forbes 1999). This often occurs as a
result of the emergence of magnetic flux into pre-existing coronal
loops, or due to a shearing of loop foot-points, with both phenomena
producing flares with different characteristics and sizes. We are not
currently able to spatially resolve stellar coronae, and therefore it
is difficult to determine the nature of stellar flares directly.
However, it is possible to get an estimate of the sizes of the
emitting volume and stellar coronal loops from the timescale of the
flare decay (Serio et al. 1991).

If we assume that the flare plasma cools predominantly by radiation
and that heating takes place at the impulsive phase of the flare with
no additional heating, we can equate the observed flare e-folding
decay time with the radiative cooling time $t_{rad}$. Serio et
al. 1991 discuss the applicablity of this first assumption, while the
latter is generally the case for implusive flares. For a flare with a
temperature $T$ the average electron density of the flare $n_e$ is
given by
\begin{equation}
 n_e = \frac{3kT}{t_{rad} P(T)}
\end{equation}
where $k$ is the Boltzmann constant and $P(T)$ is the radiative loss
function for a unit emission measure. Mewe, Gronenschild \& van den
Oord (1985) showed that for flares with temperature higher than 20MK
this function can be approximated as P(T) $\sim$
10$^{-24.73}$T$^{0.25}$erg cm$^3$s$^{-1}$. We can use the flare
temperatures which we derived from our spectral fits (see Table
\ref{fits}) to make estimates of the electron density (shown in Table
\ref{looplength}).

Based on our spectral fits we have also determined the emission measure
($EM$) for the four flares from sources with known distances. Pallavicini,
Tagliaferri \& Stella (1990) have shown that this can be expressed for a
flare composed of $n_L$ loops each with an average length $L$, as
\begin{equation}
EM= n_L {n_e^2} ( 2\pi\beta^2 L^3 )
\end{equation}
where $\beta$ is the ratio between radius of the loop cross-section
and its semi-length. Typical values of $\beta$ for Solar coronal loops
are $\sim 0.1-0.3 $ (Golub et al. 1980). We have used the above
relationship, assuming the emission originates from a single coronal
loop, to derive the loop half-lengths for the four flares which we
have presented in Table \ref{looplength}.

To understand the significance of these estimates, and draw any
conclusions about the nature of the flares on these stars it is
necessary to know the stellar radii, $R_s$.  We use the spectral type
of Source 1 and 6 as found in \S \ref{nature}, and the mass of Source
3 as $\sim$1.10{\Msun} (Pillitteri et al. 2003) to estimate the stars
radius assuming they are on the main sequence (Allen 1973). We find
that Source 1 and Source 3 both have $R_s\sim$1.05{\Rsun}, while
Source 6 is much larger $R_s\sim$5.3{\Rsun}. The length of the flare
loop estimated for Source 1 is therefore $\sim$0.7$R_s$, for Source 6
it is $\sim$0.4$R_s$, while for both flares on Source 3 the loop is of
the order of the stellar radii.

Solar flares originating from coronal loops of the order of a Solar
radii are relatively uncommon, and are generally associated with
transequatorial loop flares between different active regions (Glover
et al. 2003).  These flares are generally found to be very faint, and
are very difficult to distinguish from the average full Solar X-ray
flux. The brightest Solar flares are often very compact ($L$ $\sim$
0.05 {\Rsun}), usually being composed of multiple compact loops and
often forming arcades (Pallavicini et al. 1990). It is possible that
the large flares observed on Source 3, are in fact due to magnetic
reconnection in a combination of multiple smaller loops, rather than
one excessively large loop. If we assume average loop half-lengths of
$\sim$0.05$R_s$, this would suggest $n_L \sim20$.

\begin{table}
\begin{tabular}{cccccc}
\hline
 Source & $EM$ & $t_{rad}$ & $n_e$ & $L$ \\
        &  ($10^{53}$cm$^{-3}$) & (ks) & ($10^{11}$cm$^{-3}$) & ($10^{10}$cm)\\
\hline
1 & 8.7 & 4.6 & 4.5 & 2.6 \\
3a & 5.8 & 4.0 & 1.9 & 4.0 \\
3b & 5.3 & 3.4 & 2.0 & 3.7 \\
6 & 13.8 & 1.2 & 1.3 & 6.9 \\
\hline
\end{tabular}
\caption{Estimates for the loop half length ($L$) of the four flares
which a distance to the source was established. Also shown is the
e-folding decay time ($T_{rad}$), an estimate for the electron density
($n_e$), and a calculation of the emission measure ($EM$).}
\label{looplength}
\end{table}

\section{Investigation into the double flare}

Source 3 showed two rapid impulsive flares during the same
observation, in contrast with all the other sources which showed only
one. Flares on the Sun are observed frequently, at many wavelengths
and over many timescales, as a result of continuous
observation. Prominent Solar flares are sometimes observed within a
short time interval: they can be described in a sympathetic or
homologous way.

Sympathetic flares are thought to occur when a flare initiates a shock
front, a coronal disturbance, which propagates to another active
region on the Solar disk causing that active region to flare. These
disturbances are known as Moreton waves (Moreton 1964), and are
observed in the chromosphere typically traveling at Alfv\'enic
speeds.  The shocks are normally followed by a slower moving density
enhancement in the corona, which travels at about a third of the
Alfv\'{e}n speed (Chen et al. 2002). Another way in which active
regions may communicate is via conduction fronts which propagate along
magnetic structures linking remote regions (\v{S}vestka et al. 1977).

Homologous flares are caused by the recurrent destruction and
reformation of the stellar magnetic fields, driven by the constant
action of emerging flux and/or magnetic sheer (Martres et al.
1984). This often produces flares of similar magnitude and emission
measure which originate from the same active region.  We now
investigate whether either of these processes are occurring in Source
3.

\begin{figure}
\begin{center}
\setlength{\unitlength}{1cm}
\begin{picture}(8,4.)
\put(-0.4,0.5){\includegraphics{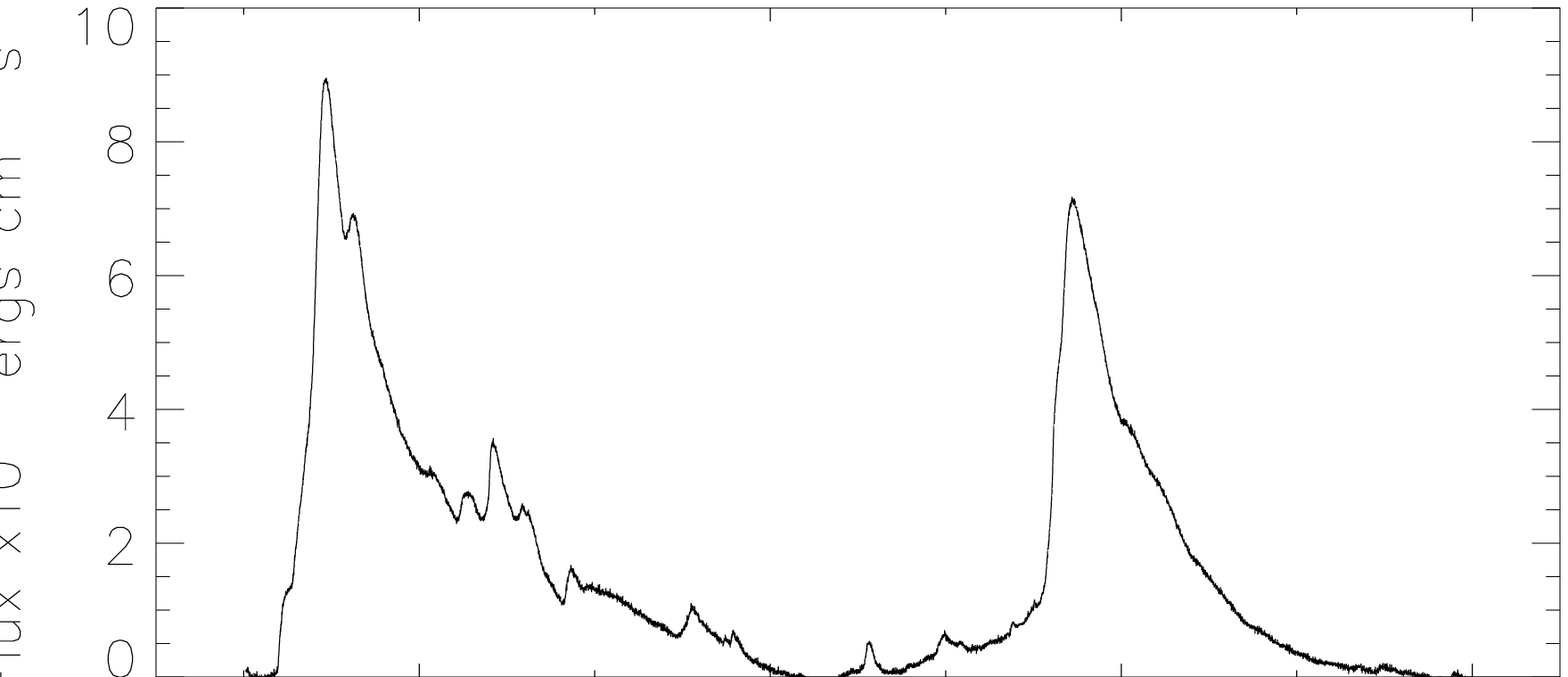}}
\end{picture}
\end{center}
\caption{A double Solar flare event in the wavelength range 1-8\AA
(1.5--12.4 keV). The flare took place on Dec 16 2001.}
\label{solarf} 
\end{figure}

\subsection{Sympathetic Activity}

Using the procedure detailed in \S \ref{Spec} we extracted a spectrum
for the quiescent state of Source 3 (assumed to be after the end of
the second flare). We found the spectrum could be fitted with a single
temperature plasma model with $T_{quiescent}=0.8$keV. Brown, Melrose
\& Spicer (1979) showed that the speed of a conduction front can be
given by the ion sound speed
\begin{equation}
 c_{i} = \sqrt \frac{kT}{m_i}
\end{equation}
where $m_i$ is the mean mass of an ion (Brown et al. 1979). Here, we
take $T$ to be $T = T_{flare} - T_{quiescent}$.  For a difference in
temperature of $\sim1.5\times10^7K$, (as determined for Source 3), a
conduction front would propagate at a speed of $\sim$320 kms$^{-1}$.

Pearce \& Harrison (1990) studied flare activity of 15 different
active regions on the Sun, and found that active region pairs
separated by less than $\sim35^{\circ}$ displayed significant amounts
of coincident flaring, while much less correlation was found between
more widely separated pairs.

Recalling our estimate of the stellar radii of Source 3, from the
previous section ($R_s\sim$1.05{\Rsun}), a conduction front travelling
at a speed of $\sim$320 kms$^{-1}$, from the first flare on Source 3
would have reached active regions with more than a $\sim35^{\circ}$
separation in just $\sim$25 mins. It would take even less time if the
propagation from the first flare was instead a Moreton wave, which
typically travel in the range 500--1000 kms$^{-1}$ on the Sun (Moreton
1964). We find that at a speed of $\sim$320 kms$^{-1}$ the conduction
front from the first flare on source 3 would have travelled across the
whole stellar disc in just over $\sim$2 hours, which is a little under
half of the delay time of 4 hours and 20 mins observed between the two
flares. Even if there were significant errors in the estimates of
stellar radius and the conduction front speed, it is highly unlikely
that the second flare is sympathetic to the first.

\subsection{Homology, a comparison with Solar flares}

We searched a section of the {\it X-ray sensor} (XRS) archives from
the {\it Geostationary Operational Environmental Satellites} (GOES)
satellite for Solar flare activity that had a similar temporal profile
and interval between flares as seen on Source 3. In Figure
\ref{solarf}, we present the lightcurve of two such Solar flares that
were observed by GOES on 2001 December 16. We used the ratio of the
counts recorded in the 1.5--12.4 keV and 3--25 keV energy bands of
GOES (Figure \ref{solarf}), to measure the average emission measure,
and electron temperature of the two Solar flares, (see Table
\ref{solar}). The derived luminosity was found to be the same in both
Solar flares: $\sim$1.3$\times10^{26}$ergs s$^{-1}$.

The two Solar flares both show a rapid rise to peak brightness with
longer decay times.  The decay time is shorter in the second of the
two flares, $\sim$130 mins as opposed to $\sim$160 mins in the first
flare. The second flare is also at a lower temperature and has a
smaller emission measure than the first flare (see table \ref{solar})
There is a delay time of $\sim$4 hours 20 mins between the peak
brightness of each flare. The characteristics mentioned above are all
consistent with those found on the double flare event observed on
Source 3.

\begin{table}
\begin{tabular}{cccccc}
\hline
 Solar & EM & T & \multicolumn{2}{c} {Luminosity}\\
 Flare &  ($10^{52}$cm$^{-3}$) & (KeV) & ($10^{26}$ergs s$^{-1}$) & ($10^{30}$ergs) \\
\hline
1 & 2.1 & 0.7 & 1.3 & 1.3 \\
2 & 1.7 & 0.6 & 1.3 & 1.1 \\
\hline
\end{tabular}
\caption{The Emission Measure, average Temperature, and luminosity for
the two Solar flares observed by GOES on December 16 2001.}
\label{solar}
\end{table}

We examined the images of the flares from the {\it Soft X-ray imager}
(also on GOES) and the {\it Extreme ultraviolet Imaging Telescope} (on
the {\it Solar and Heliospehric Observatory}). Both flares have a
similar morphology and appear to originate from the same active
region, NOAA 9733, (with a GOES classification of M1.0 and C8.5
respectively). This, coupled with the similar luminosity of each
flare, indicates that the flares are homologous in nature.

This suggests that the stellar double flares which we observed on
Source 3, may be homologous in nature and from the same stellar active
region.  The only significant difference between the stellar and Solar
case is that the stellar flares seen on Source 3 have emission
measures and temperatures larger than the flares seen on the Sun. This
could be explained by the existence of a stronger magnetic field in
the stellar active region. According to the chromospheric evaporation
model of Shibata \& Yokoyama (1999), a factor of $\sim$100 difference
in the respective magnetic field strengths could explain the larger
luminosity of the stellar flares.

\section{Conclusions}

We have searched the {\xmm} public data archive for sources showing
X-ray flares. We found 8 objects which show light curves with flare
like behaviour showing that the {\xmm} archive is an excellent
resource with which to find such objects. We extracted the EPIC PN
spectra (0.2--10keV) for each of the events and derived physical
parameters such as temperature, emission measure and flux. Using the
optical identification we estimated distances to three of the sources,
permitting calculations of the flare luminosity and an estimate of the
flare loop half-lengths. These are consistent with previous estimates
of flaring coronal loops on other stellar sources.

Source 3 showed two flares of similar magnitude and duration. We
estimate that the loop length of each flare is approximately the
length of the stellar radius. It is also possible the flare is
composed of many smaller flare loops, as is often observed on the
Sun. In this case, we estimate the number of loops to be of the order
of 20.

We have investigated the origin of this `double' flare by making an
analogy with a Solar double flare event which had a light curve
similar to the one on Source 3.  We have explored whether the process
driving these flares were sympathetic or homologous in nature.  An
estimate of the speed of a conduction front propagating over the
stellar surface, along with our estimate of the stellar radius
excludes the possibility that these stellar flares are sympathetic in
nature. On the other hand an investigation of the similar Solar event
implies it is likely that we are observing homologous flares from just
a single stellar active region on Source 3. In this scenario, the
interval observed between the flares reflects the time required for
the reconfiguration of the corona back to the original pre-flare
state, thus permitting the region to flare again under the same
conditions and resulting in a similar flare event.  The major
difference between the comparable Solar and stellar observation is the
magnitude of the flare event, which can be attributed to a stronger
stellar magnetic field of the order of a 100$\times$ the Solar
magnetic field.

\section{Acknowledgments}

The authors would like to thank Louise Harra, Sarah Matthews, Lydia
van Driel-Gesztelyi, Hilary Kay and David Williams for helpful
discussions and suggestions. DT would like to thank Mark Cropper for
helpful comments and also for supporting his work placement at the
Mullard Space Science Laboratory, enabling the work presented in this
paper to be done. This paper is based on observations obtained with
{\xmm}, an ESA science mission with instruments and contributions
directly funded by ESA Member States and the USA (NASA), and made use
of archival material from SIMBAD, which is operated at CDS,
Strasbourg. This research has also made use of the USNO-B1.0 catalogue
using the HEASARC search page. We thank the referee for helpful
comments.

\end{document}